\documentclass[twocolumn,prc,showpacs,a4]{revtex4}
\usepackage{epsfig,graphicx,float,pst-all}
\usepackage{amssymb}
\usepackage{mathrsfs}

\newcommand{\be}{\begin{equation}} 
\newcommand{\ee}{\end{equation}}
\newcommand{\bea}{\begin{eqnarray}} 
\newcommand{\eea}{\end{eqnarray}}
\newcommand{\bc}{\begin{center}}
\newcommand{\ec}{\end{center}}

\begin{document}

\title{ 
Pairing in the continuum: the quadrupole response of the Borromean nucleus $^6$He}
\author{L. Fortunato$^1$, R. Chatterjee$^2$, Jagjit Singh$^1$ A. Vitturi$^1$}
\affiliation{ 1) Dipartimento di Fisica e Astronomia ``G.Galilei''  and INFN-Sezione di Padova,\\ via Marzolo 8, 
I-35131 Padova, Italy \\
2) Department of Physics, Indian Institute of Technology, Roorkee 247 667, India}

\begin{abstract}
The ground state and low-lying continuum states of $^6$He are found within a shell model scheme, in a basis of two-particle states built out of continuum $p$-states of the unbound $^5$He nucleus, using a simple pairing contact-delta interaction. This accounts for the Borromean character of the bound ground state, revealing its composition. We investigate the quadrupole response of the system and we put our calculations into perspective with the latest experimental results. The calculated quadrupole strength distribution reproduces the narrow $2^+$ resonance, while a second wider peak is found at about 3.9 MeV above the g.s. energy.
\end{abstract}

\maketitle

Certain many-body nuclear systems stand out among light nuclei because of their peculiar constitution. They are made up of three parts, normally a core that corresponds to a stable bound nucleus plus two weakly-bound neutrons. These systems are special because if any of the three particles is removed, the resulting two-body system is unbound and it falls apart in a very short time. This is quite different from the behavior of more standard tightly-bound nuclear systems like $^{42}$Ca, or $^{210}$Pb, that sit closer to the stability valley. In these cases the removal of a single neutron does not alter the system so much as to break it and systems like $^{41}$Ca or $^{209}$Pb are perfectly bound systems. Different is the case of $^{11}$Li or $^6$He, light nuclei close to the neutron drip-lines, where the removal of a single neutron makes a big difference because neither $^{10}$Li nor $^5$He exist in bound form. Obviously the removal of the core also yields an unbound subsystem (two neutrons cannot form bound states).
The name Borromean nuclei has been coined \cite{Zh} to describe these special three-body bound systems. 

Years ago, Migdal \cite{Mi} proposed a qualitative argument in which an explanation of the stability of systems made of a core, $A$, plus two neutrons despite the intermediate system (A+n) being unbound, is given. This is linked to the presence of a resonant state in the continuum of the latter. The whole continuum in this approach is approximated with a single resonant state to which global averaged properties are attributed, discarding the specificity of the continuous spectrum.

Hansen and Jonson \cite{Ha} proposed that many of the properties of two-neutron Borromean systems can be studied and explained with a two-body model that describes the system with a core plus a dineutron cluster. It is clearly a coarse, but well-working, approximation for a correlated pair of neutrons, that interact via a NN potential. The dineutron is an idealization that would not exist alone in a bound form, but can be thought of existing {\it in medium} due to the stabilizing or  binding presence of the core's mean field.

In many instances in the last few years (for example the nice papers of Mei and van Isacker \cite{Mei}), the $0^+$ g.s. of $^6$He has been calculated by using bound, exponentially decaying, shell model states for the sake of simplicity. While this is an insightful assumption, a proper treatment of the continuum and the estimation of the role of residual interaction between single-particle continuum states is mandatory.
A standard procedure is to adjust some appropriate bound state to the energy of the resonance, couple two of them and calculate the diagonal pairing matrix elements as an integral, $\int \Psi ' V(\mid \vec r_1- \vec r_2 \mid) \Psi$ in shorthand notation, with some suitable pairing interaction, most often a contact delta interaction or a Gaussian potential.

These Borromean systems are not yet fully understood. Successful phenomenological models and {\it ab initio} models have been used to describe their structure to a reasonable degree and to approximate their behavior in nuclear reactions fairly well, but they still fail to incorporate effects due to the presence of the continuum. These are essential to understand the prime reason of their stable character. In fact these approaches normally take as a starting point for calculations not the true continuum, but rather a basis set of bound, exponentially decaying wave functions (obtained, for example, from a diagonalization in a box of finite radius). It is the purpose of this paper to show how an extension of theoretical concepts related to residual interactions, namely a contact delta pairing interaction, naturally explain the stable character of the bound states of Borromean nuclei, such as $^6$He and simultaneously account for some of the resonant structures seen in the low-lying energy continuum.  The paradigm for this type of calculations is taken from the successful calculation 
of properties of deeply-bound nuclei that have two particles outside of a doubly magic core: for example see the discussion on $^{18}$O in Heyde's textbook 
\cite{Kris}, a deeply-bound nucleus where the continuum does not play any role. The diagonal and non-diagonal matrix elements of the residual interaction give a non-trivial contribution that furnishes an utterly convincing explanation for the level structure of these nuclei.
Based on this descriptions of standard nuclear systems having two-particles outside closed shells, there have been several studies \cite{Vitt} aimed at showing how Borromean systems are bound due to the effect of pairing that brings the energy of the subsystem below the neutron emission threshold. The short-range nature of the residual NN interaction between the otherwise unbound neutrons is what kills the oscillating tail of the continuum wave functions. 
But, returning to our system, the energy of the unbound neutrons is not just a single energy, it is rather smeared on a continuous energy range according to some distribution. How does the different energies, all present at the same time with different probability, combine into a single bound state?
 
\begin{figure}[!t]
\begin{picture}(200,180)(0,-25)
\psset{unit=1.pt}\def\megaelectronvolt{20.}
\newlength\mev \mev=\megaelectronvolt\psunit
\psline[linestyle=dashed](0,0\mev)(200,0\mev)
\rput(20,160){$^5$He}  \rput(90,160){2part.} \rput(170,160){$^6$He}

\psframe[linecolor=lightgray,fillstyle=solid,fillcolor=lightgray](20,0.\mev)(30,6.84\mev)
\psframe[linecolor=lightgray,fillstyle=solid,fillcolor=lightgray](10,0.105\mev)(20,1.463\mev)
\psline{-}(10,0.789\mev)(20,0.789\mev)\rput(0,0.75\mev){$p_{3 \over 2}$}
\psline{-}(20,1.27\mev)(30,1.27\mev)\rput(0,1.30\mev){$p_{1 \over 2}$} 

\psline{-}(80,1.578\mev)(90,1.578\mev)\rput(70,1.378\mev){$p_{3 \over 2}^2$} \rput(105,1.378\mev){$0^+, 2^+$}
\psline{-}(80,2.06\mev)(90,2.06\mev)\rput(65,2.06\mev){$p_{3 \over 2}p_{1 \over 2}$} \rput(105,2.06\mev){$1^+,2^+$}
\psline{-}(80,2.54\mev)(90,2.54\mev)\rput(70,2.64\mev){$p_{1 \over 2}^2$} \rput(98,2.64\mev){$0^+$}

\psframe[linecolor=gray,fillstyle=solid,fillcolor=gray](160,0.\mev)(170,6\mev)
\pscircle[linecolor=gray,fillstyle=solid,fillcolor=gray](165,6.5\mev){3}
\pscircle[linecolor=gray,fillstyle=solid,fillcolor=gray](165,7\mev){3}
\psframe[linecolor=gray,fillstyle=solid,fillcolor=gray](150,0.703\mev)(160,0.929\mev)
\psline{-}(150,-0.973\mev)(190,-0.973\mev)\rput(148,-0.85\mev){$0^+$}
\psline{-}(150,0.816\mev)(160,0.816\mev)\rput(148,1.\mev){$2^+$}
\psline{-}(160,4.63\mev)(170,4.63\mev)\rput(135,4.65\mev){$(2^+,1^-,0^+)$}

\psframe[linecolor=pink,fillstyle=solid,fillcolor=pink](170,0.\mev)(180,3.227\mev)
\psframe[linecolor=pink,fillstyle=solid,fillcolor=pink](180,0.\mev)(190,6\mev)
\pscircle[linecolor=pink,fillstyle=solid,fillcolor=pink](185,6.5\mev){3}
\pscircle[linecolor=pink,fillstyle=solid,fillcolor=pink](185,7\mev){3}
\psline[linecolor=red]{-}(170,1.627\mev)(180,1.627\mev)\rput(176,1.9\mev){\red $2^+$}
\psline[linecolor=red]{-}(180,4.327\mev)(190,4.327\mev)\rput(200,4.4\mev){\red $1^{(\pm)}$}

\psline{->}(-10,-1.2\mev)(-10,8\mev)   \rput(-10,-1.5\mev){MeV}
\psline(-12,-1\mev)(-8,-1\mev) 
\psline(-12,0\mev)(-8,0\mev)\rput(-20,0\mev){0.0}
\psline(-12,1\mev)(-8,1\mev) 
\psline(-12,2\mev)(-8,2\mev)\rput(-20,2\mev){2.0}
\psline(-12,3\mev)(-8,3\mev) 
\psline(-12,4\mev)(-8,4\mev)\rput(-20,4\mev){4.0}
\psline(-12,5\mev)(-8,5\mev)
\psline(-12,6\mev)(-8,6\mev)\rput(-20,6\mev){6.0}
\psline(-12,7\mev)(-8,7\mev)
\end{picture}
\caption{(Color online) Left: experimental energy levels (resonances) in $^5$He. 
Center: unperturbed energies of two particle states built upon the scheme on the left. 
Right: experimental energy levels (bound ground state and resonances) of $^6$He. 
Shades of gray and pink indicate widths. The experimental energies are from Ref. \cite{TUNL} (in black) and Ref. \cite{Moug} (in red). Parentheses indicate uncertain spin-parity assignment.} \label{comp}
\end{figure}
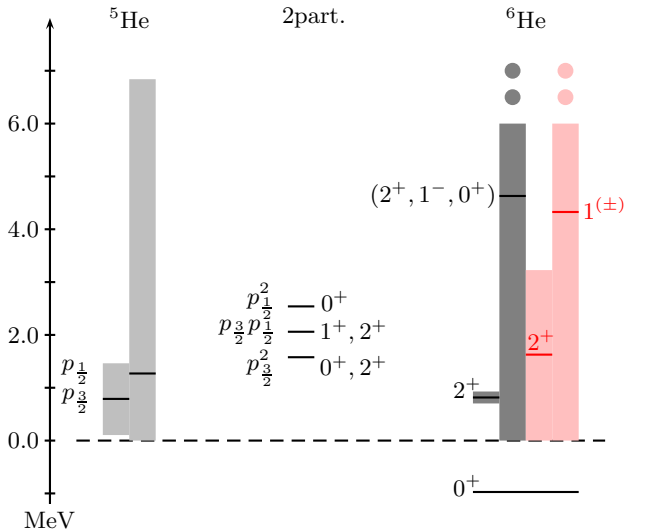

\begin{figure}[!t]
\begin{center}
\vspace{-1.cm}
\epsfig{file=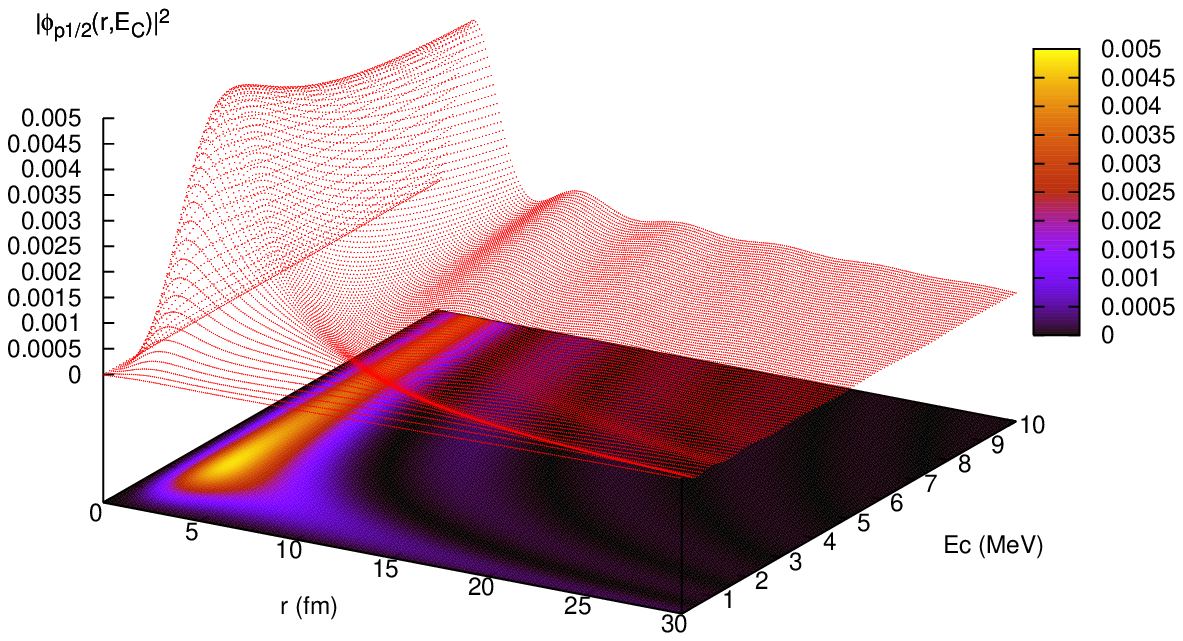,width=0.48\textwidth, bb=50 50 410 302}\\
\vspace{-1.5cm}
\epsfig{file=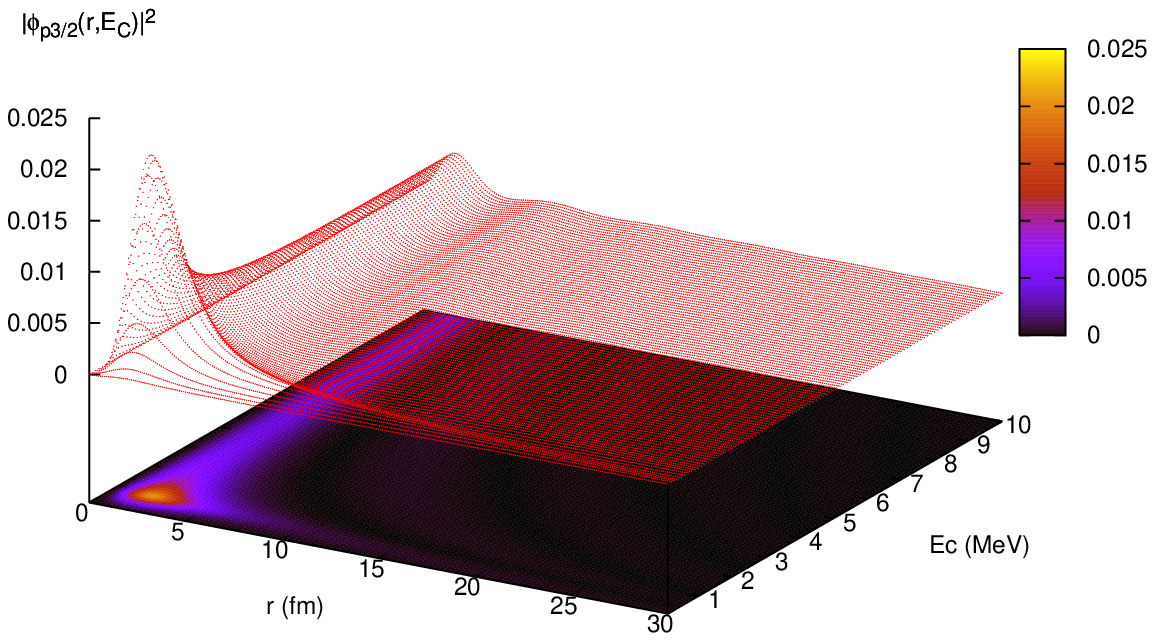,width=0.48\textwidth, bb=50 50 410 302}
\end{center}
\caption{ (Color online) Calculated square of $^5$He continuum wave functions (top: $p_{1/2}$, bottom: $p_{3/2}$) as a function of radial variable and continuum energy.} \label{waves}
\end{figure}

We start from the description of the unbound subsystem and we specialize our arguments to the lightest prototypical case of $^6$He. The subsystem $^5$He is unbound, it exists only as a short-lived resonance that breaks up into the $\alpha+n$ channel. The shell model predicts a bound, completely filled, $s$ state for the $\alpha$ core and an unbound $p$ doublet, further split by spin-orbit interaction. Experimentally the $p_{3/2}$ and $p_{1/2}$ resonances are found at 0.789 MeV and 1.27 MeV above the neutron separation threshold \cite{TUNL}. Their width are quoted as 0.648 MeV and 5.57 MeV respectively (See Fig.\ref{comp}). Note that these values have been extracted from raw data within R-matrix approach. The relative motion wave of the neutron with respect to the core is an unbound ($E_C>0, k> 0$), oscillating dipole ($\ell=1$) wave that must approximate a combination of spherical Bessel functions at large distances from the center. The continuum single-particle states of $^5$He can be reproduced fairly-well with a Woods-Saxon (WS) potential of depth $V_0=-42.6$ MeV, $r_0=1.2$ fm and $a=0.9$ fm, with a spin-orbit coefficient of $V_{ls}=8.5$ MeV (see Eq. 2-144 of Ref. \cite{BoMo}). These wavefunctions are shown in Fig. \ref{waves}. We have followed also a more refined approach, consisting in identifying the resonances through the phase shifts. In this case the width is connected to the first derivative of the phase shift. The poles of the S-matrix have been calculated with the Jost functions for WS + spin-orbit potentials. The result is that, for $V_0=-41.2$ MeV and $V_{ls}=6.5$ MeV, one gets a $p_{3/2}$ resonance with real part 0.79 and imaginary part 0.49. The $p_{1/2}$ resonance comes at 1.27  MeV with a width of 1.62 MeV. Both calculations give similar outcomes, with similar parameters, although the widths are not in perfect agreement. Therefore for the sake of easing the following calculations, we will use the simpler results coming from the first approach.

Fig. (\ref{comp}) also shows the $0^+$ ground state of $^6$He that is bound by 0.973 MeV and the $2^+$ narrow resonant state found at 1.797 MeV $\pm$25 keV above the ground state. According to standard databases another resonance is found at about 5.6 $\pm$0.3 MeV with uncertain spin-parity assignment (given as $2^+,1^-,0^+$). No other states are present up to 14 MeV. The widths of these resonances are 113 $\pm$20 keV (narrow $2^+$) and 12.1 $\pm$1.1 MeV (very broad) respectively. 
Recent experimental observations \cite{Moug}, trying to disentangle the complicated nature of the $^6$He continuum with the p($^8$He,t) reaction at SPIRAL (GANIL), support the existence of two resonances above the neutron separation energy $S_n$: a $2^+$ state at 2.6(3) MeV with $\Gamma=$1.6(4) MeV and a $1^{(+,-)}$ state at 5.3(3) MeV with $\Gamma=$2(1) MeV. These states are shown in red in Fig. (\ref{comp}). Several theories are listed in Ref. \cite{Moug} and compared with the available experimental information. Most of them show the same set of levels that we have constructed on $p$ orbitals. The somewhat puzzling nature of the excited states of $^6$He has been discussed recently in Ref. \cite{For} where it is concluded that the spin and parity of the 5.3 MeV state is most probably $0^+$, in contrast with the analysis proposed with the experimental data.

The crudest model with two non-interacting particles in the above single-particle levels of $^5$He produces 5 positive-parity states when two neutrons are  placed in the $p_{3/2}$ and $p_{1/2}$ unbound orbits. Namely the $p_{3/ 2}^2$ configuration couples to $J=0,2$ (these states can naturally be assumed as the main components of the two lowest states of $^6$He), the $p_{3/2}p_{1/ 2}$ configuration couples to $J=1,2$ and the $p_{1/2}^2$ configuration couples only to $J=0$. The unperturbed energies of these configurations are 1.578, 2.06 and 2.54 MeV respectively, as indicated in the second column of Fig.(\ref{comp}). {\it Ab initio} theories \cite{bar,Moug} (at the level of 12 $\hbar \omega$) find the sequence of levels ($0^+, 2^+, 2^+, 1^+, 0^+$) with a third $0^+$ lowering rapidly as the basis is increased (see Fig.1 of cited Ref. \cite{bar}), confirming our simpler scheme.
A few relevant statements can now be done: if the $1^-$ attribution of part of the strength is confirmed, then {\it its nature cannot be associated with 2 neutrons sitting in $p$ orbitals}.
One might wonder whether dipole strength should be present: certainly a highly collective dipole mode built at high excitation energy as a coherent superposition of p-h states must exist, but its nature in $^6$He calls for promotion of one neutron from the p to the sd shell. The low-energy tail of this giant dipole resonance indeed is expected to come down till zero and therefore it might mix significantly with other states and make the spin-parity assignment very difficult.
This strength might also come from other configurations, such as $\alpha+(2n)$ cluster configurations, that mix with the $\alpha+n+n$ configuration.

\begin{figure}[!t]
\begin{center}
\epsfig{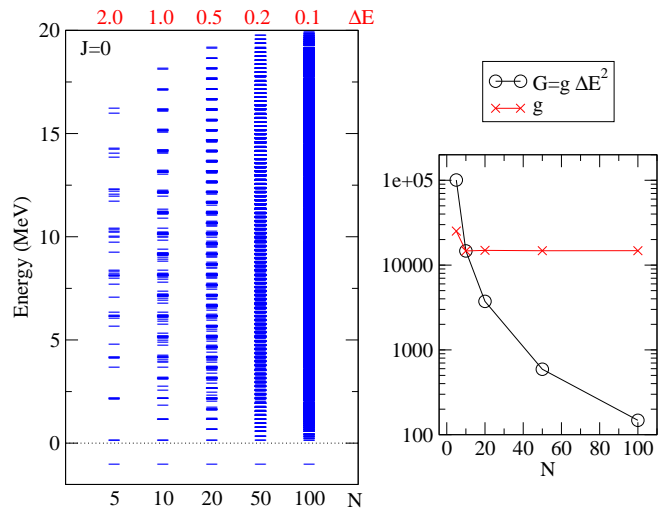}
\end{center}
\caption{(Color online)  Left: Eigenspectrum of the interacting two-particle case for $J=0$ for increasing basis dimensions, $N$. The coefficient of the $\delta-$contact matrix, $G$, has been adjusted each time to reproduce the g.s. energy (right). The actual strength of the pairing interaction, $g$, is obtained by correcting with the energy spacing $\Delta E$ and it is practically a constant.} 
\label{eig-vs-basis}
\end{figure}

The five states discussed above are not discrete, but rather depend on the energies of the two continuum orbitals $$\phi_{\ell,j,m}(\vec r, E_C)=\phi_{\ell,j}(r, E_C)[Y_{\ell m_\ell}(\Omega)\times \chi_{1/2,m_s}]^{(j)}_m \;.$$  These can be combined into a tensor product two-particle wave function, 
$$
\psi_{JM}(\vec r_1, \vec r_2)=[\phi_{\ell_1,j_1,m_1}(\vec r_1,{E_C}_1) \times \phi_{\ell_2,j_2,m_2}(\vec r_2, {E_C}_2)]^{(J)}_M $$
that must be discretized and used as a basis for calculations.
It is clear that one needs at least to introduce the residual interaction between continuum states, a task that requires careful numerical implementation because one deals with large datasets. We take an attractive pairing contact delta interaction, $-g\delta(\vec r_1 - \vec r_2)$ for simplicity, although, as it is well-known, density dependent interactions might be more appropriate \cite{BerEs}. We have calculated the continuum single-particle wavefunctions, with energies from 0.0 to 10.0 MeV, normalized to a delta similarly to Ref. \cite{Aust}, for the p-states of $^5$He on a radial grid that goes from 0.1 fm to 100.0 fm with the potential discussed above (Notice that this amount to 2.4 Gb of data for each component). With these wavefunctions, using the mid-point method with an energy spacing of 2.0, 1.0, 0.5, 0.2 and 0.1 MeV, corresponding to block basis dimensions of $N=$5, 10, 20, 50 and 100 respectively, we formed the two particle states and calculated the matrix elements of the pairing interaction ($\sim$ 4Gb of data for the largest case). This has been diagonalized with standard routines and it has given the eigenvalues shown in Fig. (\ref{eig-vs-basis}) for the $J=0$ case. The coefficient of the $\delta-$contact matrix, $G$, has been adjusted to reproduce the correct ground state energy each time. The actual pairing interaction $g$ is obtained by correcting with a factor that depends on the aforementioned spacing between energy states and it is practically a constant, except for the smallest basis.  The biggest adopted basis size gives a fairly dense continuum in the region of interest.

\begin{figure}[!t]
\begin{center}
\epsfig{file=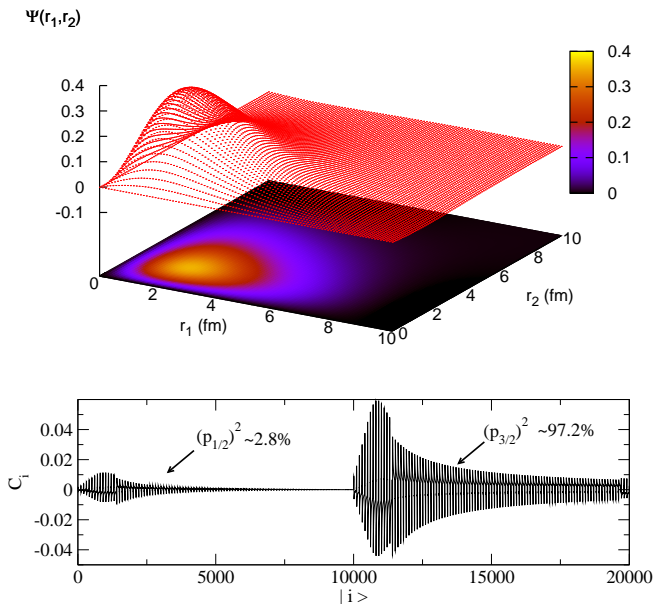 ,width=0.48\textwidth , clip=, bb= 50 50 320 215}
\epsfig{file=compo-graph-ampli.eps ,width=0.48\textwidth , clip=}
\end{center}
\caption{(Color online) Ground state wavefunction ($S=0$) for $N=$100 as a function of the coordinates of the two neutrons and corresponding contour plot (upper part). Decomposition of the g.s. into the J=0 basis (lower part) as a function of an arbitrary basis state label: the basis is divided in two blocks, $10^4$ $[p_{1/2}\times p_{1/2}]^{(0)}$ components and then $10^4$ $[p_{3/2}\times p_{3/2}]^{(0)}$ components. The ordering in each block is established by the sequential energies of each pair of continuum s.p. states, i.e. $(E_{C_1},E_{C_2}) =$ (0.1, 0.1), (0.1, 0.2), $\dots$ ,(0.1, 10.0), (0.2, 0.1), (0.2, 0.2), $\dots$ (10.0, 10.0).} 
\label{groundsts}
\end{figure}

The radial part of the $S=0$ g.s. wavefunction obtained from the diagonalization in the largest basis is displayed in the upper part of Fig.(\ref{groundsts}). 
Due to symmetry reasons and to the fact that $\ell_1=\ell_2=1$, there is no $S=1$ component for a $\delta-$interaction (see Ref. \cite{ShTa}, ch. 20). In fact, in this case, we can write the two-particle wavefunction as $\Psi(\vec r_1, \vec r_2)= \Psi(r_1, r_2) \mathscr{Y}^+_{JM}(\Omega_1,\Omega_2) \chi_{S=0}$.
It is symmetric with respect to the exchange of coordinates of the two identical neutrons. It shows a certain degree of collectivity, taking contributions of comparable magnitude (though not all of the same sign) from several basis states, while in contrast the remaining unbound states usually are made up of a few major components. The surface plot shows the exponential behavior typical of a bound state, despite being the sum of many products of oscillating wavefunctions.
One can see from the bottom part of the figure that the square of the amplitudes of the $(p_{3/2})^2$ components are dominant summing up to 97.2\%. 
In principle also the s-continuum should be introduced in the picture, because the $(s_{1/2})^2$ configuration of course couples to $J=0$.We did not introduce it in the calculation because the p-resonances dominate the $^5$He spectrum and because the numerical computations are already very demanding. Following Ref. \cite{HagSa}, however, the contribution of $p^2$ in $^6$He is estimated to be the most relevant with a percentage of about 83\%. In the lower part of the Fig.(\ref{groundsts}) one can also see that the chosen cut in energy is appropriate because the states with a label approaching $10^4$ and $2\cdot 10^4$ become progressively less and less important.

Notice that, in our approach, there is no information on the angular correlation, that has nevertheless been extensively investigated by various authors: it corresponds to the $\tilde \mathscr{C}_{\ell^2;00}(\theta_{12})$ of Ref. \cite{Mei}.

\begin{figure}[!t]
\begin{center}
\epsfig{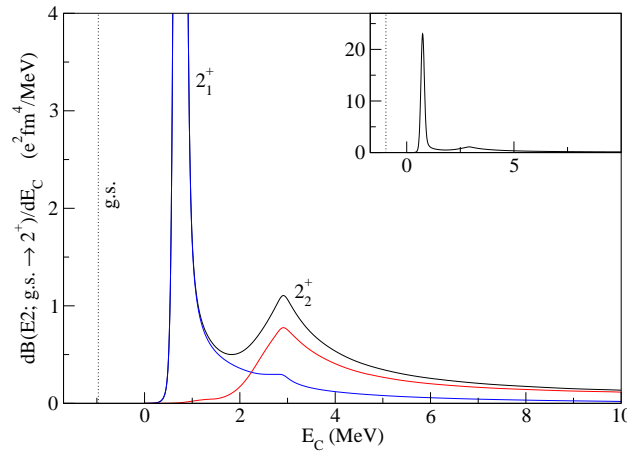}
\caption{(Color online) Quadrupole strength distribution with respect to the break up threshold. The total strength (black) is split into the contribution of the  $(p_{3/2})^2$ and $(p_{3/2}p_{1/2})$ components, in blue and red respectively. The insert shows the full curve for the total strength.} 
\label{be2}
\end{center}
\end{figure}

While most theoretical studies have focused on dipole strength \cite{Aoya, Desc}, we have performed a set of calculations for quadrupole transitions.
After constructing a basis of the same size made up of two parts, namely $[p_{3/2}\times p_{3/2} ]^{(2)}$ and $[p_{3/2}\times p_{1/2} ]^{(2)}$, 
we diagonalize the pairing matrix and obtain eigenvalues, that are all unbound, and the corresponding eigenvectors. 

We compute the $B(E2)$ values (Fig.\ref{be2}) to these eigenvalues and adjust the strength of the pairing matrix to get the energy centroid of the first peak at about the right position ($E=0.76$ MeV, $\Gamma \sim 0.2$ MeV). The width is a bit larger then the experimental value. We also obtain a second peak at about $E=2.91$ MeV with an asymmetric width at half maximum of $\Gamma \sim 1.8$ MeV. While the first peak is mainly due to $(p_{3/2})^2$ components, the second peak is clearly identified as arising mainly from $(p_{3/2}p_{1/2})$ components. Measuring energies from the g.s., the second peak is found at about $3.88$ MeV. A noteworthy feature of this peak is that it is found at an energy higher than the corresponding unperturbed two-particle state, despite the attractive nature of pairing: this is a consequence of the asymmetric long tail in energy of the $p_{1/2}$ resonance in $^5$He. 
The total integrated strength amounts to about $8.8$ $e^2fm^4$, of which about 3/4 is in the first peak. This value can be compared with the
value of 9.7471 $e^2fm^4$ obtained in Ref. \cite{Lay}. To the best of our knowledge Fig. 6 of the cited paper is the only published theoretical prediction for E2 strength distribution in $^6$He and with some little differences, we essentially confirm that result.  
According to our calculations the second peak does not match with the recently identified $2^+$ strength at 2.6 MeV \cite{Moug}. Possibly other components, like $s$ and $d$ continuum states of $^5$He, when taken into account theoretically might affect the quadrupole response of $^6$He. We plan to thoroughly investigate this aspect. Several theories disagree on the predictions for the low-lying continuum states of $^6$He and the available experimental information cannot be considered as completely free from model assumptions. For example in Ref.\cite{Moug} one might wonder that the procedure of extracting the position and width of the secondary $2^+$ state from the shoulder of the primary $2^+$ resonance peak is not free from arbitrariness in the choice of background. This is all the more true when dealing with exotic beams with low intensity and when several competing processes might create a background, as in the present case. The choice of Breit-Wigner parameterization is another factor might also influence the outcomes. Certainly a lager body of experiments is needed in order to unravel the structure of low-lying resonances of $^6$He.

We have shown how the bound Borromean ground state of $^6$He emerges from the coupling of two unbound p-waves in the $^5$He continuum, due to the presence of the pairing interaction. Other similar studies have used artificially bound p-states or have used a box to discretize the continuum. We analyze the E2 response showing where we expect two resonances to occur. The stark mismatch between theory and experiments on the position of the higher resonance calls for further work, because our understanding of drip-line Borromean systems passes through the proper description of the lightest and foremost example of them, $^6$He. 

We would like to thank J.A.Lay and P.Descouvemont for useful suggestions. J.Singh gratefully acknowledges the finanacial support from Fondazione Cassa di Risparmio di Padova e Rovigo (CARIPARO).

\end{document}